\documentclass{eas}
\usepackage{graphicx}
\newcommand{\salt}[6]{{#1}$_{#2}${#3}$_{#4}${#5}$_{#6}$}

\usepackage[normal]{subfigure}
\usepackage{float}
\begin{document}

%%-----------------------------
%%      the top matter
%%-----------------------------

\title{Carbonate Formation in Non-Aqueous Environments by Solid-Gas Carbonation of Silicates} 
\author{S.J. Day}\address{Astrophysics Group, Keele University, Staffordshire, ST5 5BG, UK}
\secondaddress{Diamond Light Source, Harwell Campus, Didcot, Oxon, OX11 0DE, UK}
\author{S.P. Thompson}\sameaddress{2}
\author{A. Evans}\sameaddress{1}
\author{J.E. Parker}\sameaddress{2}

\runningtitle{Day \etal: Carbonate Formation in Astrophysical Environments}
\begin{abstract}
We have produced synthetic analogues of cosmic silicates using the Sol Gel
method, producing amorphous silicates of composition \salt{Mg}{(x)}{Ca}{(1-x)}{SiO}{3}. 
Using synchrotron X-ray powder diffraction on Beamline I11 at the Diamond Light Source, together with a
newly-commissioned gas cell, real-time powder diffraction scans have been taken
of a range of silicates exposed to CO$_{2}$ under non-ambient conditions. The SXPD is complemented by other techniques including Raman and Infrared Spectroscopy and SEM imaging.
\end{abstract}
\maketitle

%%-----------------------------
%%      your text
%%-----------------------------
\vspace{-10pt}
\section{Introduction}

Dust is present throughout the universe, existing in a
wide variety of astrophysical environments and playing a
major role in star and planet formation. Throughout their
lifetime, dust grains are exposed to high levels of radiation
and extreme temperatures, leading to the evolution of the
amorphous, structurally disordered grains, that are characteristic of the diffuse interstellar medium (ISM; \cite{kemper04}), to the highly processed materials that were present in the early solar nebula. We are investigating this processing
and mineralization of grains using laboratory produced analogues of cosmic dust. In-situ synchrotron X-ray powder diffraction (SXPD)
is conducted on Beamline I11 at Diamond Light Source (Thompson \textit{et al.} 2009, 2011) and
is complemented by SEM imaging, Raman and FTIR spectroscopy. This has relevance
to the existence of carbonates in circumstellar environments that are
believed to form through solid-gas interaction of cosmic
silicates with gaseous CO$_2$ at high temperatures (\cite{toppani05,rietmeijer08}). It also has relevance to processes in planetary environments where there is abundant CO$_2$ in the gas phase (\cite{johnsonfegley02}).

We describe preliminary proof-of-concept work which demonstrates that this process can indeed proceed.

\section{Experimental}
\vspace{-2pt}
Beamline I11 is equipped to simulate extreme
temperature and gas pressure conditions (20 $<$
$T$ $<$ 1000$^{\circ}$C, 1-100 bar gas pressure), through
the use of a cyber\-star hot air blower and gas
cell capillary holder (\cite{parker2012}; Thompson \textit{et al.} these proceedings). Amorphous magnesium and \-calcium silicates of
composition \salt{Mg}{x}{Ca}{1-x}{SiO}{3}, where 0 $<$ $x$ $<$ 1, were
produced in the laboratory as analogues of cosmic dust,
using an adapted sol gel method (\cite{thompson2012}). To inhibit the direct formation of
carbonate, the gels were dried \textit{in vacuo} and the
dried silicates stored under argon. Powdered samples were loaded
into quartz capillaries in a gas cell and
mounted onto the sample stage of the diffractometer.
The heating and dosing of samples with CO$_2$ was
controlled remotely allowing real time, in-situ measurements to be
taken. SXPD data were collected using a fast Position Sensitive Detector (\cite{thompson2011}),
specifically designed for fast data collection. Data were
collected at regular intervals as the samples were put under vacuum and then exposed to CO$_2$ up to a pressure of 1 bar. While this is clearly higher than in circumstellar environments (so the reaction rate is correspondingly higher), the nature of the experimental setup requires such pressure in order to provide adequate diffusion of CO$_2$ through the high packing density of the material in the capillary. Further, the experimental CO$_2$ pressure is comparable with that which was likely to be present in early planetary atmospheres (\cite{wood68}).
After a number
of scans at room temperature the hot air blower was introduced and the
temperature was increased in steps of 50$^{\circ}$C, over a period of $\sim1$ hour, up to a maximum of 900$^{\circ}$C; two 4-s scans were taken at each temperature.
\vspace{-10pt}
\begin{center}
\begin{figure}[h]
  \begin{center}
%  \subfigure[Dried in
% air]{\label{semair}\includegraphics[scale=0.15]{SEM_Air_Dried.jpg}} 
% \hspace{45pt}
\subfigure[SEM image, sample dried in vacuum]{\label{semvac}\includegraphics[width=50mm, height=35mm]{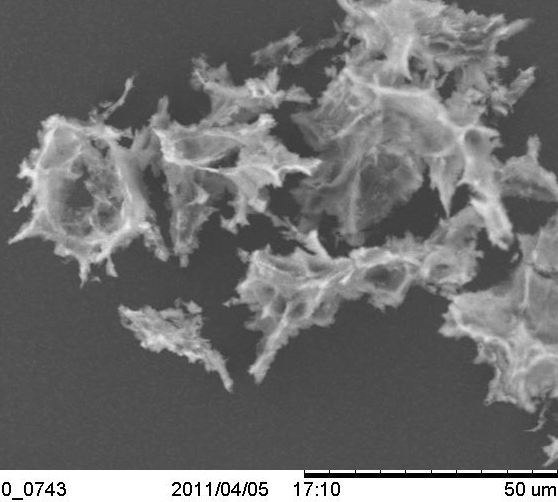}}
\subfigure[Raman Spectra]{\label{raman}\includegraphics[width=62mm, height=37mm]{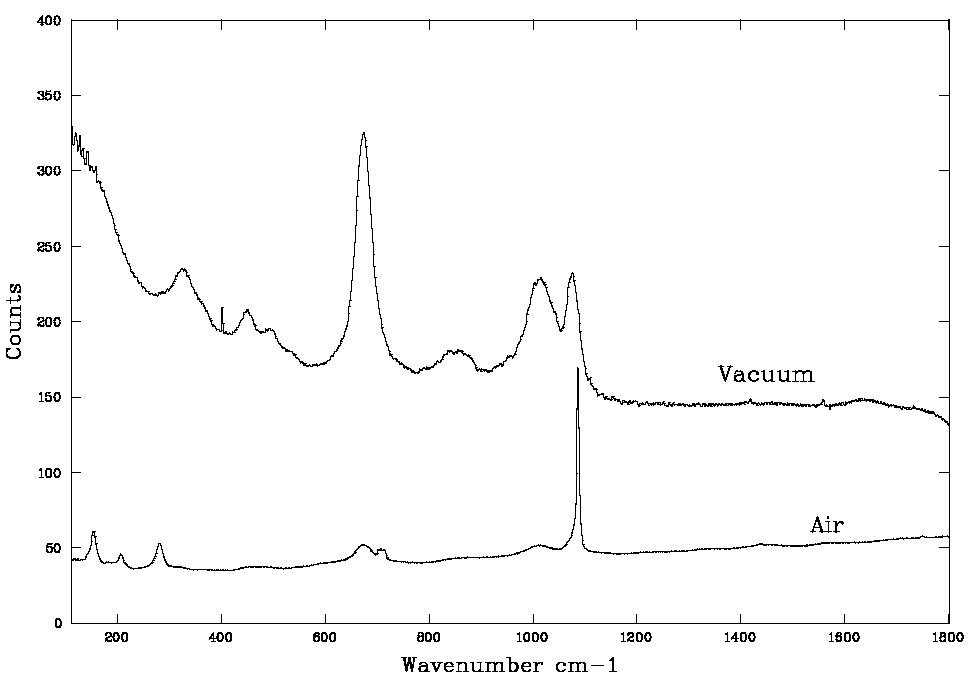}} \\
  \caption{SEM image of vacuum dried sample and a comparison of Raman Spectra of MgCaSiO$_3$ grains, dried in air and under vacuum. \label{compare}}
  \end{center}
\end{figure}\end{center}

\section{Characterization}

\subsection{Raman Spectroscopy}

The sol gels were dried in air and \textit{in vacuo}, producing noticeably different end products. The gel dried in air produced solid, angular grains of a few mm to a few cm in size, varying from clear to white in colour depending on composition. On the other hand, gels dried \textit{in vacuo} produce a very fine white powder, with SEM images showing a branch-like network structure with individual grains no larger than a few $\mu$m in size (see Figure~\ref{semvac}).

\begin{figure}[h]
  \begin{center}
 \subfigure[Compositional]{\label{FTIRcomp}\includegraphics[width=60mm, height=45mm]{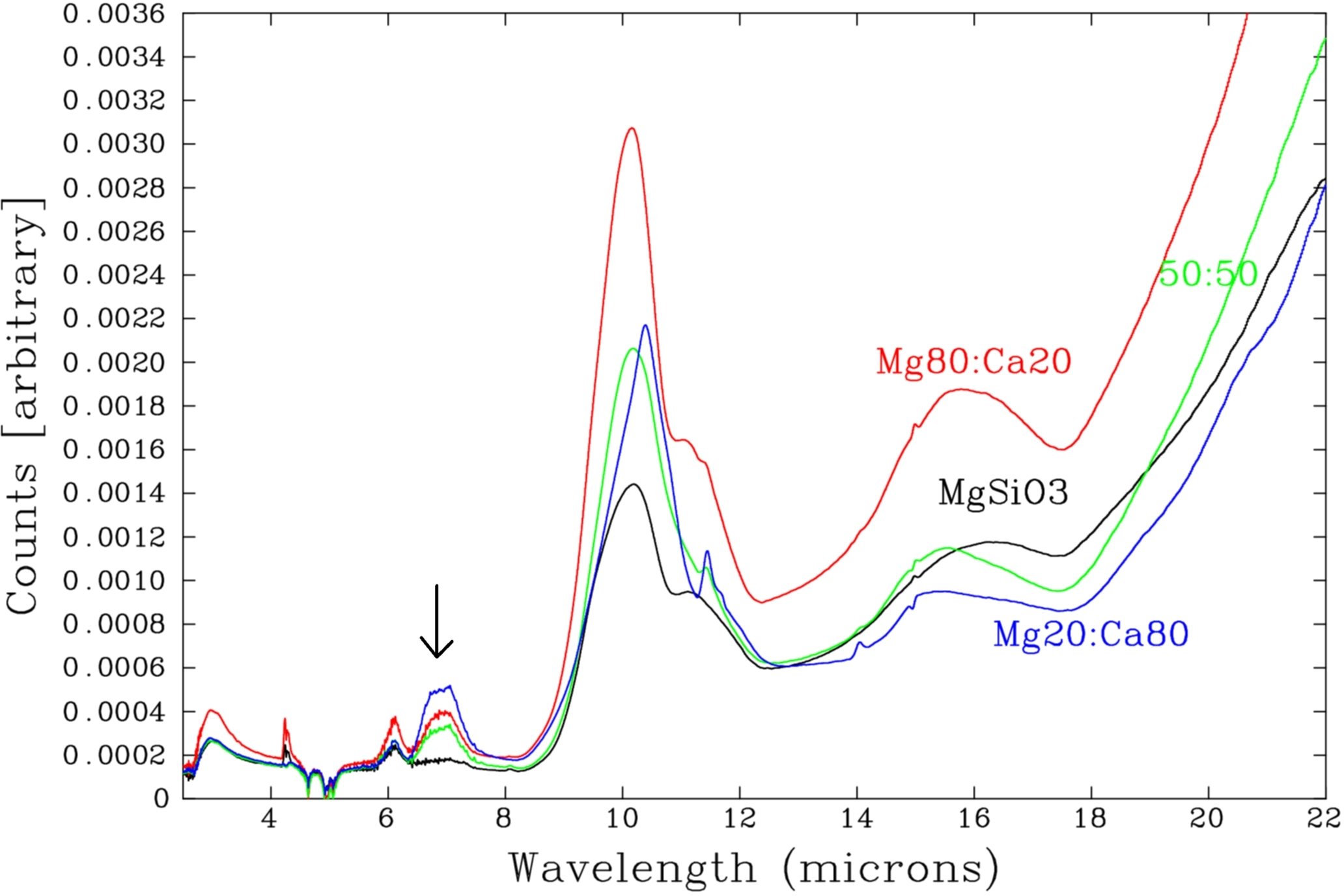} }
\subfigure[Thermal]{\label{FTIRtemp}\includegraphics[width=61mm, height=45mm]{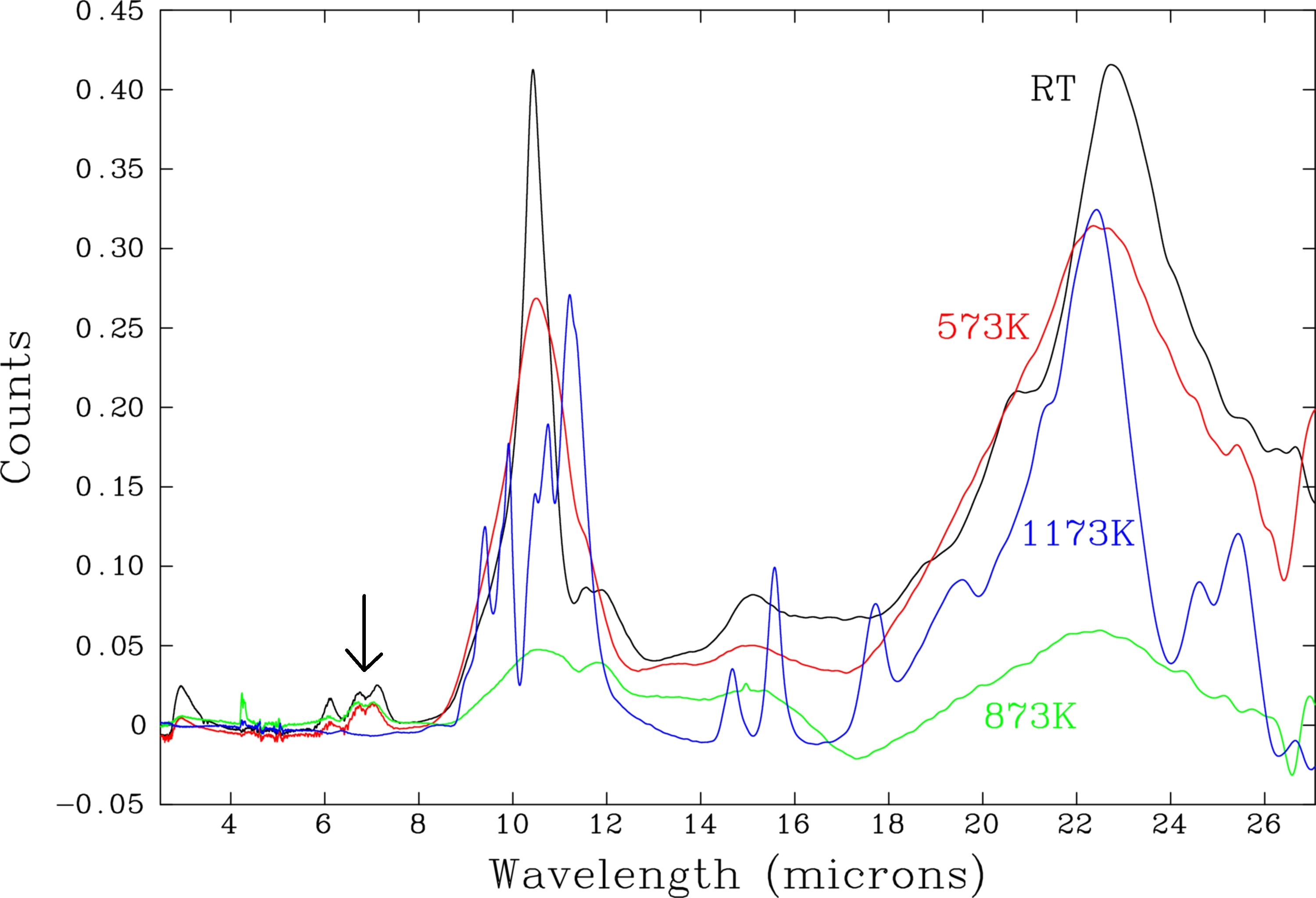}}
  \caption{Comparison of FTIR Spectra of MgCaSiO$_3$ grains, of varying composition and temperature. \label{ftircompare}}
  \end{center}
\end{figure}

\vspace{-10pt}
Raman spectra show that there is also a compositional difference between the samples dried in air and those dried \textit{in vacuo} (see Figure~\ref{raman}). The air-dried samples exhibit features of crystalline calcium carbonate (calcite) exhibiting characteristic features at 712cm$^{-1}$ and 1088cm$^{-1}$, while samples dried \textit{in vacuo} show a predominately silicate composition with prominent features in the 670cm$^{-1}$ and 1000cm$^{-1}$ bands. This implies that the carbonation of the samples is due to the reaction of Ca with atmospheric CO$_2$ during the drying phase.

\begin{center}
\begin{figure}[h]
\begin{center}
\subfigure[Crystallization sequence.]{\label{cryst} \includegraphics[scale=0.25]{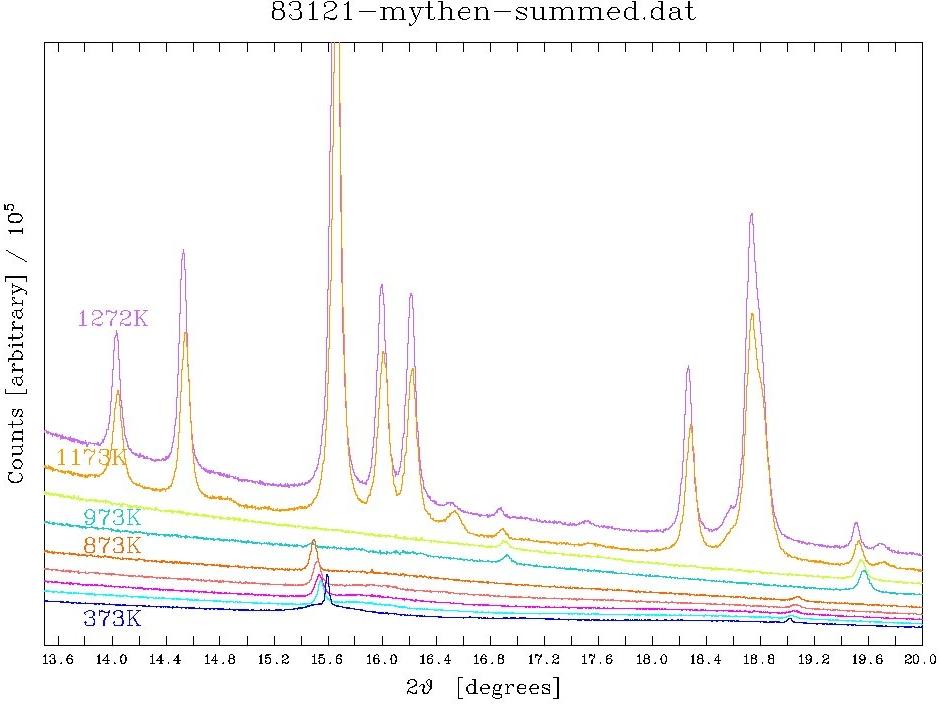}}
\subfigure[Weight Percentages of Diopside and Calcite]{\label{wtperc} \includegraphics[width=60mm, height=38mm]{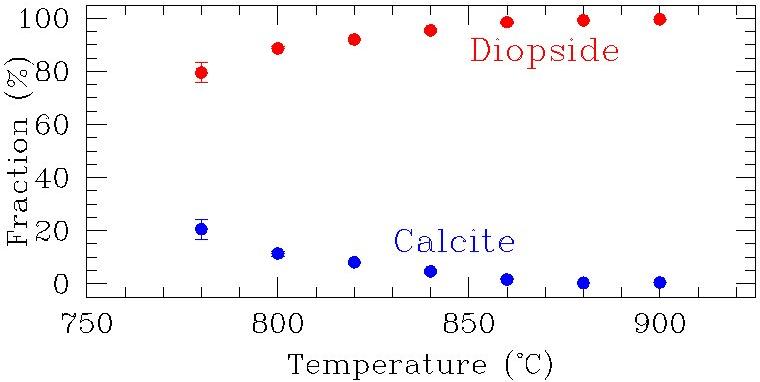}}                         \end{center}
\caption{\label{reit} Sequence of powder diffraction patterns showing the structural evolution of a silicate sample and derived weight percentages of diopside and calcite phases.}
\end{figure}
\end{center}

\subsection{FTIR}

FTIR spectra of the samples were taken for a range of compositions and annealing temperatures.
Figure~\ref{FTIRcomp} shows the comparison of four compositions, $x = 0, 0.2, 0.8$ and $1$. A peak (arrowed) can be clearly seen at 7 $\mu$m
strengthening with increasing Ca content. Also with increasing Ca content, the broad silicate features around 11$\mu$m and 16$\mu$m seem to weaken. Figure~\ref{FTIRtemp} shows the spectra of a CaSiO$_3$ sample annealed at four different temperatures. Few differences are observed at temperatures of up to 873K, but note the disappearance of the 7$\mu$m feature and the rise of crystalline silicate features at temperatures greater than 873K. This indicates that any carbonate phase present within the sample becomes unstable at this temperature and breaks down.

\subsection{Synchrotron X-Ray Powder Diffraction}

 Figure~\ref{cryst} shows a sequence of in-situ SXPD patterns of \salt{Mg}{0.5}{Ca}{0.5}{SiO}{3} exposed to CO$_2$ at 1 bar pressure and heated to 1272K over a period of a few hours. A feature indicative of calcite is present at low temperatures but disappears above 873K, again suggesting that the calcite phase is not stable at higher temperatures. The sample then begins to crystallize to diopside at just below 1173K. Rietveld refinement\footnote{Rietveld refinement involves the fitting of a model crystal structure to x-ray powder diffraction data in order to determine the mineral phases present within the sample and to obtain information about the crystal structure.} of the powder diffraction patterns, using the TOPAS-Academic software package (\cite{coelho07}), has provided information about the relative weight percentages of the diopside and calcite phases within the sample as it is being heated. This shows an inverse correlation of the two phases (see Figure~\ref{wtperc}).
\vspace{-5pt}

\section{Results and Conclusions}

Analysis of these data is still ongoing and the crystallization
temperatures and effect of CO$_2$ on the samples will be
much better constrained through further experiments and analysis. At this stage we provide proof of physical principle regarding the formation of carbonates rather than a direct simulation of any astrophysical environment. However
by knowing that the carbonate phase becomes unstable
above 873K, we can begin to constrain the astrophysical
environments (both circumstellar and planetary) in which carbonates would be able to form
and survive.
\vspace{-10pt}
%%-----------------------------
%%      your bibliography
%%-----------------------------

\end{document}